\documentclass[runningheads]{svmult}
\usepackage{epsfig}
\usepackage{makeidx}   
\usepackage{graphicx}  
\usepackage{subeqnar}  
\usepackage{multicol}  
\usepackage{physprbb}  
\makeindex             

\begin{document}
\title*{The Hamiltonian Mean Field Model:
from Dynamics to Statistical Mechanics and back}


\titlerunning{The Hamiltonian Mean Field Model}

\author{Thierry Dauxois\inst{1}
\and Vito Latora\inst{2} \and Andrea Rapisarda\inst{2} \and
Stefano Ruffo\inst{1,3}
 \and Alessandro Torcini\inst{3,4}}

\authorrunning{T. Dauxois et al.}
\institute{Laboratoire de Physique, UMR CNRS 5672, ENS Lyon, 46,
  all{\'e}e d'Italie, F-69007 Lyon, France \and Dipartimento di Fisica,
  Universit{\'a} di Catania, and Istituto Nazionale di Fisica Nucleare,
  Sezione di Catania, Corso Italia 57, I-95129 Catania, Italy \and
  Dipartimento di Energetica "S. Stecco", Universit{\`a} di Firenze, via
  S. Marta, 3, INFM and INFN, I-50139 Firenze, Italy \and
  UMR-CNRS 6171,  Universit{\'e} d'Aix-Marseille III -
  Av. Esc. Normandie-Niemen, F-13397 Marseille Cedex 20, France.}

\maketitle              

\begin{abstract}
The thermodynamics and the dynamics of particle systems with
infinite-range coupling display several unusual and
new features with respect to systems with short-range
interactions. The Hamiltonian Mean Field (HMF) model
represents a paradigmatic example of this class of systems.
The present study addresses both attractive and repulsive
interactions, with a particular emphasis on the description
of clustering phenomena from a thermodynamical as well as
from a dynamical point of view. The observed
clustering transition can be first or
second order, in the usual thermodynamical sense. In the former
case, ensemble inequivalence naturally arises close to the
transition, i.e. canonical and microcanonical ensembles give
different results. In particular, in the microcanonical ensemble
negative specific heat regimes and temperature jumps are
observed. 
Moreover, having access to dynamics one can study non-equilibrium
processes. Among them, the most striking is
the emergence of coherent structures in the repulsive model,
whose formation and dynamics can be studied either by using the tools of
statistical mechanics or as a manifestation of the
solutions of an associated Vlasov equation. 
The chaotic character of the HMF model has been also analyzed in terms
of its Lyapunov spectrum.
\end{abstract}

\section{Introduction}
Long-range interactions appear in the domains of
gravity~\cite{paddyhmf,chavanishmf} and of plasma physics~\cite{elskenshmf} and
make the statistical treatment extremely complex.  Additional features
are present in such systems at short distances: the
gravitational potential is singular at the origin and screening
phenomena mask the Coulomb singularity in a plasma. This justifies the
introduction of simplified toy models that retain only the long-range
properties of the force, allowing a detailed description of the
statistical and dynamical behaviors associated to this feature. 
In this context a special role is
played by mean-field models, i.e. models where all particles interact
with the same strength. This constitutes a dramatic reduction of
complexity, since in such models the spatial coordinates have no role,
since each particle is equivalent.  However, there are several
preliminary indications that behaviors found in mean-field models
extend to cases where the two-body potential decays at large distances
with a power smaller than space dimension~\cite{tamahmf,BMRhmf,campa}.

The
Blume-Emery-Griffiths (BEG) mean-field model is discussed in
this book~\cite{BMRhmf} and represents an excellent benchmark to discuss
relations between canonical and microcanonical ensembles.  Indeed,
this model is exactly solvable in both ensembles and is, at the same
time, sufficiently rich to display such interesting features as
negative specific heat and temperature jumps in the microcanonical
ensemble. Since these effects cannot be present in the canonical
ensemble, this rigorously proves ensemble inequivalence. However, the
BEG model has no dynamics and only the thermodynamical behavior can be
investigated. Moreover, it is a spin model where variables take discrete
values. It would therefore be amenable to introduce a model that
displays all these interesting thermodynamical effects, but for which
one would also dispose of an Hamiltonian dynamics with continuous
variables, whose equilibrium states could be studied both in the
canonical and in the microcanonical ensemble. Having access to
dynamics, one could moreover study non equilibrium features and
aspects of the microscopic behavior like sensitivity to initial
conditions, expressed by the Lyapunov spectrum~\cite{Ruelleeckman}.
Such a model has been introduced in Ref.~\cite{Antoni} and has been
called the Hamiltonian Mean Field (HMF) model. In the simpler version,
it represents a system of particles moving on a circle, all coupled by
an equal strength attractive or repulsive cosine interaction.  An
extension of it to the case in which particles move on a 2D torus has
been introduced in Ref.~\cite{at} and it has been quite recently
realized that all such models are particular cases of a more general
Hamiltonian~\cite{art}. 

The HMF model, that we introduce in Section 2,
is exactly solvable in the canonical ensemble by a
Hubbard-Stratonovich transformation. The solution in the
microcanonical ensemble can be obtained only under certain hypotheses
that we will discuss in Section 3, but detailed information on the
behavior in the microcanonical ensemble can be obtained by direct
molecular dynamics (MD) simulations. The model has first and second
order phase transitions and tricritical points. Its rich phase diagram
allows to test the presence of ensemble inequivalence
\index{ensemble inequivalence} near canonical first order phase transitions and,
indeed, we find negative specific heat and temperature jumps in the
microcanonical ensemble. Having access to dynamics, one can study
metastability of out-of-equilibrium states. This is done in Section 4,
where we analyze the emergence of a coherent structure in the
repulsive HMF at low energy.  Similar features are also discussed in
another chapter of this book~\cite{tsallisrap} for the attractive case
near the second order phase transition.  Section 5 is devoted to the
study of the spectrum of Lyapunov exponents.  The maximal exponent has
a peak near the phase transition~\cite{at,lat1hmf,celiahmf} and vanishes
when increasing the number of particles with a universal scaling law
in the whole high energy disordered phase.  In a low energy range the
Lyapunov spectrum has a thermodynamic limit distribution similar to
the one observed for systems with short-range
interaction~\cite{livipolruffo}.

\section{The HMF models}

A generic two-body potential in a two dimensional square box of
side $2\pi$ with periodic boundary conditions, a 2D torus, can be Fourier
expanded as
\begin{equation}
\label{fourierexpan}
V(x,y)=\sum_{{\bf k}=(k_x,k_y)} \exp \left( i {\bf k} \cdot {\bf r}
\right) V(k_x,k_y)\quad.
\end{equation}
A sufficiently rich family of potential functions is obtained if,
we restrict to the first two momentum shells $|k|=1$ and
$|k|=\sqrt{2}$, we require that the potential is only invariant
under discrete rotations by all multiples of $\pi/4$, and we assume
that the Fourier coefficients on each shell are the same.
This amounts to perform a truncation in the Fourier expansion
of the potential (\ref{fourierexpan}), as done in studies of
spherically symmetric gravitational systems in another chapter
of this book~\cite{grossdh}.
We get
\begin{equation}
\label{limitexpansion}
V(x,y)=a+b\cos x +b\cos y +c\cos x\cos y\;.
\end{equation}
As the constant $a$ is arbitrary and scaling $b$ is equivalent to
scale the energy, $c$ remains the only free parameter.  We consider
$N$ particles interacting through the two-body potential $V(x,y)$ and
we adopt the Kac prescription~\cite{Kachmf}\index{Kac prescription}
\index{extensivity} to scale the equal strength coupling among the
particles by their number $N$.  This scaling allows to perform safely
the thermodynamic limit\index{thermodynamic limit}, since both the
kinetic and the potential energy increase proportional to
$N$\footnote{Kac prescription is, however, unphysical and it would be
  important to find a viable alternative.}.  By appropriately
redefining the constants $a=2\varepsilon+A,b=-\varepsilon,c=-A$ in formula
(\ref{limitexpansion}) and using Kac prescription one gets the
following potential energy
\begin{eqnarray}
\label{eqHA}
V_A &=& \frac{1}{2N} \sum^{N}_{i,j=1}
\varepsilon \left( 1- \cos ( x_i-x_j ) \right)
+\varepsilon \left( 1-\cos ( y_i-y_j ) \right) \nonumber \\
&& \hskip 2truecm+A \left( 1 - \cos ( x_i-x_j ) \cos ( y_i-y_j ) \right) ~,
\end{eqnarray}
with $(x_{i},y_{i})\in ]-\pi, \pi ] \times ]-\pi, \pi ]$ representing the coordinates
of $i$-th particle and $(p_{x,i},p_{y,i}) \) its conjugated momentum.
The Hamiltonian of the HMF model is now the sum of this potential
energy with the kinetic energy
\begin{equation}
\label{kinetic}
K= \sum _{i=1}^{N}\left(\frac{p_{x,i}^{2}+p_{y,i}^{2}}{2}\right)~.
\end{equation}
We get
\begin{equation}
\label{hamHMF}
H_{HMF}=K+V_A~.
\end{equation}
In the following, we will consider model (\ref{hamHMF}) for
$A=0$ and both $\varepsilon$ positive (attractive case) and
negative (repulsive case). The $A\neq0$ case will always have
$A>0$ and $\varepsilon=1$.

\section{Equilibrium Thermodynamics}

In this section, we discuss the equilibrium thermodynamical
results for model~(\ref{eqHA}) in the canonical and
microcanonical ensembles. Canonical results will be obtained
analytically while, for the microcanonical ones, we will mostly rely on
molecular dynamics (MD) simulations.

\subsection{Canonical Ensemble for $A=0$}\index{canonical!ensemble}

\label{Aeq0}
 For pedagogical reasons, we will initially limit our
analysis to the case $A=0$, for which the model reduces to two
identical {\em uncoupled} systems: one describing the evolution of
the $\{ x_i,p_{x,i} \}$ variables and the other $\{y_i,p_{y,i} \}$. 
Therefore let us rewrite the Hamiltonian
associated to one of these two sets of variables, named
$\theta_i$ in the following. We obtain
\begin{equation}
H_0 = \sum_{i=1}^N \frac{p_i^2}{2} + \frac{\varepsilon}{2N}
\sum_{i,j=1}^N [1-\cos(\theta_i - \theta_j)] = K_0 + V_0
\label{model0}
\end{equation}
where $\theta_i \in [-\pi;\pi[$ and $p_i$ are the corresponding
momenta. This model can be seen as representing particles moving
on the unit circle, or as classical $XY$-rotors with infinite
range couplings. For $\varepsilon>0$, particles attract each other
and rotors tend to align (ferromagnetic case), while for
$\varepsilon<0$, particles repel each other and spins tend to
anti-align (antiferromagnetic case). At short distances, we can
either think that particles cross each other or that they collide
elastically since they have the same mass.

The physical meaning of this model is even clearer if one
introduces the mean field vector
\begin{equation}
{\bf M} = M {\rm e}^{i \phi} = \frac{1}{N} \sum_{i=1}^N {\bf m}_i
\label{m0}
\end{equation}
where ${\bf m}_i =(\cos \theta_i, \sin \theta_i)$. $M$ and $\phi$
represent the modulus and the phase of the order parameter, which
specifies the degree of clustering in the particle interpretation,
while it is the {\em magnetization} for the $XY$ rotors. Employing
this quantity, the potential energy can be rewritten as a sum of
single particle potentials $v_i$
\begin{equation}
V_0 = \frac{1}{2} \sum_{i=1}^N v_i  \qquad {\rm with} \qquad v_i =
1- M \cos(\theta_i -\phi) \quad . \label{v0}
\end{equation}
It should be noticed that the motion of each particle is coupled to all 
the others, since the mean-field
variables $M$ and $\phi$ are determined at each time $t$ by the
instantaneous positions of all particles.

\begin{figure}[ht]
\begin{center}
\includegraphics[width=0.9\textwidth]{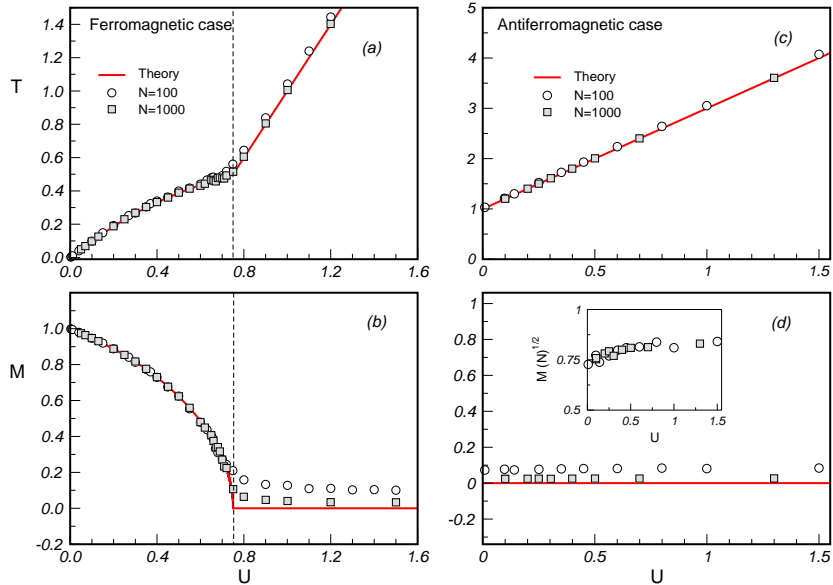}
\end{center}
\caption[]{Temperature and magnetization as a function of the
energy per particle $U$ for $A=0$ and $|\varepsilon|=1$ in the
ferromagnetic (a),(b) and in the antiferromagnetic case (c),(d).
Symbols refer to MD data for $N=10^2$ and $10^3$, while the solid
lines refer to the canonical prediction obtained analytically in
the mean-field limit. The vertical dashed line indicates the
critical energy in the ferromagnetic model, located at $U_c=0.75$,
$\beta_c=2$. The inset of panel
(d) shows the rescaled magnetization $M\times \sqrt{N}$.}
\label{fe_anti}
\end{figure}

The equilibrium results in the canonical ensemble can be obtained from
the evaluation of the partition function
\index{canonical!partition function}
\begin{equation}
Z = \int d^N p_i d^N \theta_i \exp{(-\beta H)}
\label{z0}
\end{equation}
where $\beta = 1/(k_B T)$, with $k_B$ the Boltzmann constant and
$T$ the temperature. The integration domain is extended
to the whole phase space. Integrating over momenta, one gets:
\begin{equation}
Z = \left( \frac{2 \pi}{\beta} \right)^{N/2} \int_{-\pi}^\pi d^N
\theta_i \exp{\left[\frac{-\beta \varepsilon N}{2} (1-{\bf
M}^2)\right]}\quad. \label{zz0}
\end{equation}
In order to evaluate this integral,  we use the two dimensional
Gaussian identity
\begin{equation}
\exp{\left[\frac{\mu}{2} \bf x^2\right]}=
\frac{1}{\pi} \int_{-\infty}^\infty  \int_{-\infty}^\infty
d{\bf y} \exp{[-{\bf y}^2+\sqrt{2 \mu} {\bf x}
\cdot {\bf y}]}
\label{gau}
\end{equation}
where ${\bf x}$ and ${\bf y}$ are two-dimensional vectors and
$\mu$ is positive. We can therefore rewrite Eq.~(\ref{zz0}) as
\begin{equation}
Z = \left( \frac{2 \pi}{\beta} \right)^{N/2}
\exp{\left[\frac{-\beta \varepsilon N}{2}\right]} J
\end{equation}
with
\begin{equation}
J= \frac{1}{\pi}
\int_{-\pi}^\pi d^N \theta_i
\int_{-\infty}^\infty  \int_{-\infty}^\infty
d{\bf y} \exp{[-{\bf y}^2+\sqrt{2 \mu} {\bf M}
\cdot {\bf y}]}
\label{j0}
\end{equation}
and $\mu = \beta \varepsilon N$. We use now definition~(\ref{m0}) and
exchange the order of the integrals in (\ref{j0}), factorizing the
integration over the coordinates of the particles. Introducing
the rescaled variable $ {\bf y} \to {\bf y} \sqrt{N/2\beta
\varepsilon}$, one ends up with the following expression for $J$
\begin{equation}
J= \frac{N}{2 \pi \beta \varepsilon} \int_{-\infty}^\infty
\int_{-\infty}^\infty d{\bf y} \exp{\left[-N \left( \frac{y^2}{2
\beta \varepsilon} -\ln\left(2\pi I_0(y)\right) \right) \right] }
\label{j1}
\end{equation}
where $I_n$ is the modified Bessel function of order $n$ and $y$
is the modulus of ${\bf y}$. Finally, integral~(\ref{j1}) can be
evaluated by employing the saddle point technique in the
mean-field limit (i.e. for $N \to \infty$). In this limit, the
Helmholtz free energy
 per particle $f$ reads as :
\begin{equation}
\beta f=-\lim_{N \to \infty} \frac{\ln Z}{N}=
-\frac{1}{2}\ln\left(\frac{2 \pi}{\beta}\right) +\frac{\varepsilon
\beta }{2} +\max_y\left(\frac{y^2}{2\beta\varepsilon}-\ln(2\pi
I_0(y))\right)\quad. \label{freehmf}
\end{equation}
The maximum condition leads to the consistency equation
\begin{equation}
\frac{y}{\beta \varepsilon}=\frac{I_1(y)}{I_0(y)}\quad.
\label{cons}
\end{equation}
For $\varepsilon<0$, there is a unique solution $\bar y=0$, which
means that the order parameter remains zero and there is no phase
transition (see Figs.~\ref{fe_anti}c,d). The particles are all the
time homogeneously distributed on the circle and the rotors have
zero magnetization. On the contrary, in the ferromagnetic case
($\varepsilon>0$), the solution  $\bar y=0$ is unstable for
$\beta \geq \beta_c = 2$. At $\beta=\beta_c$, two stable symmetric 
solutions appear through a pitchfork bifurcation and a discontinuity 
in the second derivative of the free energy is present, indicating a 
second order phase transition\index{phase!transition}.

These results are confirmed by an analysis of the order
parameter\footnote{This is obtained by adding to the Hamiltonian
an external field and taking the derivative of the free energy
with respect to this field, evaluated at zero field.}
\begin{equation}
M= \frac{I_1(\bar y)}{I_0(\bar y)} \quad. \label{mag0}
\end{equation}
For $\varepsilon >0$, the magnetization $M$ vanishes continuously
at $\beta_c$ (see Fig.~\ref{fe_anti}a,b), while it is always 
identical to zero in the
antiferromagnetic case (see Fig.~\ref{fe_anti}c,d). Since $M$
measures the degree of clustering of the particles, we have for
$\varepsilon >0$ a transition from a clustered phase when
$\beta>\beta_c$ to a homogeneous phase when $\beta<\beta_c$. We
can obtain also the energy per particle
\begin{equation}
U= \frac{\partial(\beta f)}{\partial \beta}=
\frac{1}{2\beta}+\frac{\varepsilon}{2}\left(1-M^2\right)
\label{u0}
\end{equation}
which is reported for $|\varepsilon|=1$ in Fig. \ref{fe_anti}.
Panels (c) and (d) of Fig.~\ref{fe_anti} are limited to the range $U>0$
because in the antiferromagnetic model a non-homogeneous state, a
{\em bicluster}, can be generated for smaller energies. The
emergence of this state modifies all thermodynamical and dynamical
features as will be discussed in section~\ref{bicluster}.

The dynamics of each particle obeys the following
pendulum equation of motion
\begin{equation}
\ddot \theta_i= -M \sin(\theta_i-\phi) \quad , \label{pend}
\end{equation}
where $M$ and $\phi$ have a non trivial time dependence, related
to the motion of all the other particles in the system.
Equation~(\ref{pend}) has been very successfully used to describe
several features of particle motion, like for instance trapping
and untrapping mechanisms~\cite{Antoni}. There are also numerical
indications~\cite{Antoni} and preliminary theoretical
speculations~\cite{priv} that taking the mean-field limit before
the infinite time limit, the time-dependence would disappear and
the modulus and the phase of the magnetization become constant.
This implies, as we will discuss in Section 5, that chaotic
motion would disappear.
The inversion of these two limits is also discussed in the
contribution by Tsallis et al~\cite{tsallisrap}.

\subsection{Canonical ensemble for $A\neq0$}\index{canonical!ensemble}

As soon as $A > 0$, the evolution along the two spatial directions
is no more decoupled and the system cannot be described in terms
of a single order parameter. Throughout all this section
$\epsilon=1$. A complete description of the phase
diagram of the system requires now the introduction of two
distinct order parameters:
\begin{equation}
{\bf M}_z= \left(\frac{\sum_i \cos z_i}{N},\frac{\sum_i \sin z_i}{N}\right)=M_z
\exp{(i \phi_z)} \label{mz}
\end{equation}
where $z_i=x_i$ or $y_i$ and;
\begin{equation}
{\bf P}_{x \pm y}= \left(\frac{\sum_i \cos (x_i \pm y_i)}{N},
\frac{\sum_i \sin (x_i \pm y_i)}{N}\right)=P_z
\exp{(i \psi_z)}\quad. \label{pz}
\end{equation}
It can be shown that on average $M_x \simeq M_y \simeq M$ 
and $P_{x+y} \simeq P_{x-y} \simeq P$: therefore, we
are left with only two order parameters.

Following the approach of section~\ref{Aeq0}, the canonical
equilibrium properties can be derived analytically in the
mean-field limit~\cite{at}. We obtain
\begin{equation}
\beta f=\frac{M^2+P^2}{\beta} -\ln
\left[\frac{G(M,P;A)}{\beta} \right]
\label{free2d}
\end{equation}
with
\begin{equation}
G= \int_0^{2\pi} ds \enskip I_0\left(M+\sqrt{2A} P \cos s\right)
\exp{(M\cos s)} \label{g2d}
\end{equation}
where $s$ is an integration variable.

\begin{figure}[ht]
\begin{center}

\includegraphics[width=\textwidth]{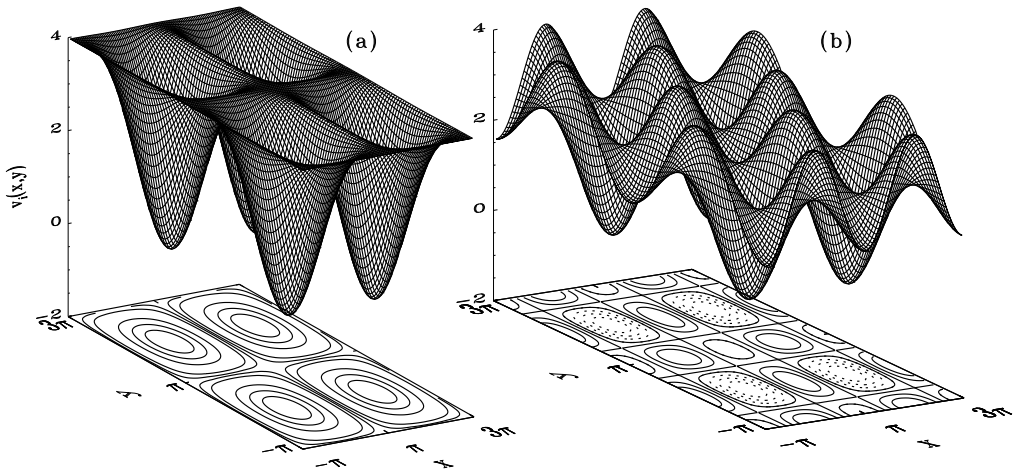}
\end{center}
\caption[]{Single particle potential $v_i(x,y)$ for $A=1,\varepsilon=1$ (a) and
$A=4,\varepsilon=1$(b).} 
\label{egg}
\end{figure}

The energy per particle  reads as
\begin{equation}
U = \frac{1}{\beta}+\frac{2+A-2 M^2-A P^2}{2}=K+V_A \quad .
\label{u2d}
\end{equation}
Depending on the value of the coupling constant $A$, the 
single particle potential defined through
$V_A = \frac{1}{2} \sum_i v_i$  changes its shape,
inducing the different clustering phenomena described below. 
For small values of $A$, $v_i$ exhibits a single minimum
per cell $(x,y)\in ([-\pi,\pi],[-\pi,\pi])$ (see Fig.~\ref{egg}a
for $A=1$), while for larger values of $A$ four minima can
coexist in a single cell (see Fig.~\ref{egg}b for $A=4$).
In the former case only one clustered equilibrium state can
exist at low temperatures, while in the latter case,
when all the four minima have the same depth a phase with
two clusters can emerge, as described in the following.

\begin{figure}[ht]
\begin{center}
\includegraphics[width=0.6\textwidth,angle=270]{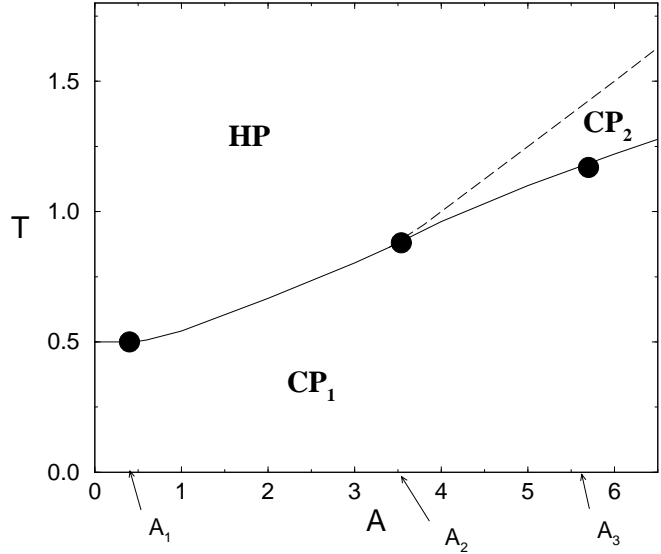}
\end{center}
\caption[]{Canonical phase diagram of model~(\ref{eqHA}) reporting the
transition temperatures versus the coupling parameter $A$. The
solid (resp. dashed) lines indicate the $T_M$ (resp. $T_P$) lines.
The dots the points where the nature of the transitions change. \protect\(
A_{1}\protect \), \protect\( A_{2}\protect \) and \protect\(
A_{3}\protect \) are the threshold coupling constants that
determine the transition scenario \protect\( I\to IV\protect \). }
\label{pda}
\end{figure}

Since the system is ruled by two different order parameters $M$
and $P$,  the phase diagram is more complicated than in the $A=0$
case and we observe two distinct clustered phases. In the very low
temperature regime, the system is in the clustered phase $
CP_{1}$: the particles have all the same location in a single
point-like cluster and $M\approx P \approx 1 $. In the very large
temperature range, the system is in a homogeneous phase ($HP$)
with particles uniformly distributed, $M\approx P=O(1/\sqrt{N})$.
For $A > A_2 \sim 3.5$, an intermediate two-clusters phase $CP_{2}
$ appears. In this phase, due to the symmetric location of the two
clusters in a cell, $M \sim O(1/\sqrt{N})$ while $P \sim O(1)$ \cite{art}.
We can gain good insights on the transitions by considering the
line $T_M$ (resp. $T_P$) where $M$ (resp. $P$) vanishes and  the
phase $CP_1$ (resp. $CP_2$) looses its stability (see Fig.
\ref{pda} for more details).

We can therefore identify the following four different scenarios
depending on the value of $A$.

\textbf{(I)} When $ 0 \leq A \leq A_1 = 2/5$, one observes a continuous
transition from the phase $CP_1$ to $HP$.  The critical line is
located at $T_M=1/2$ ($U_M=3/2+A$) and a canonical tricritical point,
located at $A_1=2/5$, separates the 1st order from the second order phase
transition\index{phase!transition} regions\footnote{
The tricritical point has been identified by finding first the
value of ${\bar P}(M,A,\beta)$ that minimizes $f(M,P;A,\beta)$ and then 
substituting the solution $\bar P$ in $f$. This reduces $f$ to a
function of $M$ only. Now, the standard
procedure described also for the BEG model \cite{BMRhmf} can be applied.
This consists in finding the value of $\beta$ and $A$ where both
the $M^2$ and the $M^4$ coefficient of the development of the free
energy in powers of $M$ vanishes
}

\textbf{(II)} When \( A_{1}<A<A_{2}\approx 3.5 \), the transition
between \( CP_{1} \) and \( HP \) is first order with a finite
energy jump (latent heat). Inside this range of values of $A$ one
also finds microcanonical discontinuous transitions (temperature
jumps) as we will see in Section 3.2.

\textbf{(III)} When \(A_{2}<A<A_{3}\approx 5.7 \), the third phase
begins to play a role and two successive transitions are observed:
first $CP_1$ disappears at $T_M$ via a first order transition that
gives rise to $CP_2$; then this two-clusters phase gives rise to the
$HP$ phase via a continuous transition. The critical line
associated to this transition is $T_P =A/4$ ($U_P =3A/4+1$).

\textbf{(IV)}  When \( A>A_{3}\),   the transition connecting the
two clustered phases, $CP_1$ and $CP_2$, becomes second order.

\subsection{Microcanonical Ensemble}\index{microcanonical!ensemble}
\label{microens}

As we have anticipated, our microcanonical results have been mostly
obtained via MD simulations, since we cannot estimate easily the
microcanonical entropy $S$ analytically for the HMF model\footnote{It
  would be possible using large deviation
  technique~\cite{barrethesis}.}. However, we show below how far we
can get, starting from the knowledge of the canonical free energy,
using Legendre transform or inverse Laplace transform techniques.  The
following derivation is indeed valid in general, it does not refer to
any specific microscopic model.

The relation that links the partition function $Z(\beta,N)$ to
the microcanonical phase-space density at energy $E=U \cdot N$
\begin{equation}
\omega(E,N) = \int d^N p_i d^N \theta_i \; \delta(E-H)
\label{w0}
\end{equation}
is given by
\begin{equation}
Z(\beta,N)=\int_0^\infty dE \;\omega(E,N) \; {\rm e}^{-\beta E}~,
\label{z0_w}
\end{equation}
where the lower limit of the integral ($E=0$) corresponds
to the energy of the ground state of the model.
Expression (\ref{z0_w}) can be readily rewritten as
\begin{equation}
Z(\beta,N)=N \int_0^\infty dU
\exp{\left[N(-\beta U + \frac{1}{N} \ln(\omega(E,N))
\right]}~,
\label{z0_w2}
\end{equation}
which is evaluated by employing the saddle-point technique in the
mean-field limit. Employing the definition of entropy per particle
in the thermodynamic limit\index{thermodynamic limit}
\begin{equation}
S(U)=\lim_{N\to \infty} \left[\frac{1}{N} \ln \omega(U,N)
\right]\quad, \label{e0}
\end{equation}
one can obtain the Legendre transform
that relates the free energy to the entropy:
\begin{equation}
-\beta F(\beta) = \max_U[-\beta U +S(U)] \qquad {\rm with} \qquad
\beta=\frac{\partial S}{\partial U} \label{f0_e0}\quad.
\end{equation}
Since a direct analytical  evaluation of the entropy of the HMF model in the
microcanonical ensemble is not possible, we are rather interested
in obtaining the entropy from the free energy. This can be done
only if the entropy $S$ is  a concave function of the energy.
Then, one can invert (\ref{f0_e0}), getting
\begin{equation}
S(U)=\min_{\beta>0} [\beta(U - F(\beta))] \qquad {\rm with} \qquad
U=\frac{\partial (\beta F)}{\partial \beta} \quad . \label{e0_f0}
\end{equation}
However, the assumption that $S$ is concave is not true for systems
with long range interactions near a canonical first order
transition, where a "convex intruder"\index{convex intruder} of $S$
appears~\cite{BMRhmf,grossdh}, which gives rise to a negative specific
heat\index{negative!specific heat} regime in the microcanonical
ensemble\footnote{A convex intruder is present also for short range
interactions in {\em finite} systems, but the entropy regains its
concave character in the thermodynamic limit~\cite{grossdh}.}.  
For such cases, we have to rely on MD simulations or microcanonical 
Monte-Carlo simulations~\cite{grossbook}.

An alternative approach to the calculation of the entropy consists in
expressing the Dirac $\delta$ function in Eq.~(\ref{w0})
by a Laplace transform. One obtains
\begin{equation}\label{invLapla}
\omega(E,N) =\frac{1}{2i\pi}\int_{-i\infty}^{+i\infty} d\beta\;
e^{\beta E}\; Z(\beta)
\end{equation}
where one notices that $\beta$ is imaginary. As the partition function can
be estimated for our model, we would just have to analytically
continue to complex values of~$\beta$. By performing a rotation to the
real axis of the integration contour, one could then evaluate $\omega$ by
saddle point techniques.  However, this rotation requires the
assumption that no singularity is present out of the real axis: this
is not in general guaranteed. It has been checked numerically for
the $A=0$ model~\cite{barrethesis}, allowing to obtain then an explicit
expression for $\omega$ and confirming ensemble equivalence for
this case.

However, when $A> A_1$, a first order canonical transition occurs,
which implies the presence of a convex intruder and makes the
evaluation of $S(U)$ through Eq.~\ref{e0_f0} impossible
\index{ensemble inequivalence}.  For these cases, a canonical 
description is unable
to capture all the features associated with the phase transition. For
the microcanonical entropy, we have to rely here on MD simulations.

The MD simulations have been performed adopting  extremely
accurate symplectic integration schemes~\cite{mac}, with
relative energy conservations during the runs of order $\sim
10^{-6}$. It is important to mention  that the CPU time required
by our integration schemes, due to the mean-field nature of the
model, increases linearly with the number of  particles.

Whenever we observe canonically continuous transitions (i.e. for
$A \leq A_1$), the MD results coincide with those obtained
analytically in the canonical ensemble, as shown in
Fig.~\ref{fe_anti} for $A=0$. The curve $T(U)$ is thus well
reproduced from MD data, apart from finite $N$ effects. It has
been however observed that starting from  "water-bags" initial
conditions metastable states can occur in the proximity of the
transition~\cite{tsallisrap}.

\begin{figure}[ht]
\begin{center}
\includegraphics[width=0.7\textwidth,angle=270]{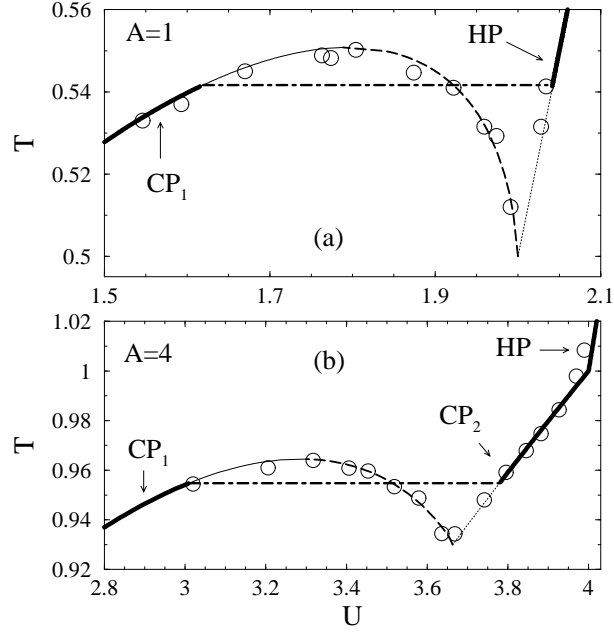}
\end{center}
\caption[]{Temperature-energy relation in the coexistence region
for A=1 (a) and A=4 (b). Lines indicate canonical analytical
results, while circles correspond to microcanonical MD
simulations. Solid thick lines are equilibrium results, solid thin
lines metastable states and dashed thin lines unstable states. The
 dash-dotted  line is the Maxwell construction. Figure (a)
refers to a first order transition from $CP_1$ to $HP$, (b) to
discontinuous transition connecting the two clustered phases. In
(b) the second order transition from $CP_2$ to $HP$ associated to
the vanishing of $P$ is also shown. The MD results refer to model
(\ref{eqHA}) with $N=5000$ averaged over a time $t=10^6$.} \label{UvT}
\end{figure}

When $A > A_1$, discrepancies between the results obtained in the two
ensembles are observable in Fig.~\ref{UvT}. For example, MD
results in the case $A=1$, reported in Fig.~\ref{UvT}(a), differ
clearly from canonical ones around the transition, exhibiting a regime
characterized by a negative specific heat\index{negative!specific heat}. 
This feature is common to many models with
long-range~\cite{thir,compa} or power-law decaying
interactions~\cite{tamahmf} as well as to finite systems with
short-range forces~\cite{grossbook}. However, only recently a
characterization of all possible microcanonical transitions associated
to canonically first order ones has been initiated~\cite{BMRhmf,art}.

\begin{figure}[ht]
\begin{center}
\includegraphics[width=0.5\textwidth,angle=270]{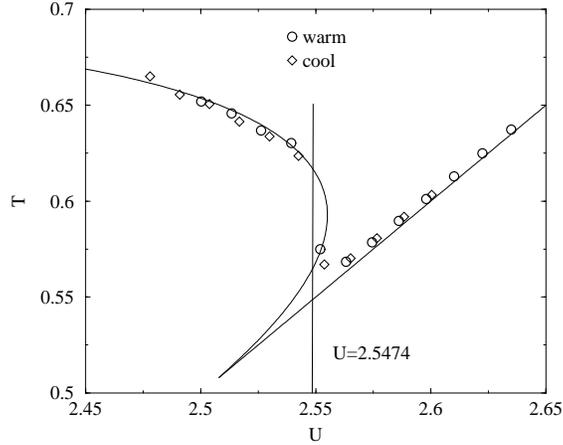}
\end{center}
\caption[]{Evolution of the temperature $T$ versus the energy $U$
in the case $A=2$. The symbols refer to MD results obtained by
successively cooling or warming a certain initial configuration.
Each simulation has been performed at constant total
energy and refers to a system of $N=4,000$ particles
integrated for a time $t=10^6$. The solid lines are computed
in the canonical ensemble and include also unstable and metastable
cases. The solid vertical line
indicating the transition energy has been estimated via
a Maxwell construction performed in the microcanonical ensemble
(for details see \protect\cite{BMRhmf}).}
\label{mdis}
\end{figure}

For $A$ slightly above $A_1$, the transition is microcanonically
continuous, i.e. there is no discontinuity in the $T-U$ relation
(this regime presumably extends up to $A\sim1.2$). Before the
transition, one observes a negative specific heat regime (see
Fig.~\ref{UvT}a). In addition, as already observed for the
Blume-Emery-Griffiths model~\cite{BMRhmf},  microcanonically {\em
discontinuous} transitions can be observed in the "convex
intruder" region. This means that, at the transition energy, {\em
temperature jumps} exist in the thermodynamic limit. A complete
physical understanding  of this phenomenon, which has also been
found in gravitational systems~\cite{chavanishmf}, has not been
reached. For $A> 1.2$, i.e.  above the "microcanonical tricritical
point", our model displays temperature jumps. This situations is
shown in Fig. \ref{mdis} for $A=2$.

For $A>A_2$, we have again a continuous transition connecting the
two clustered phases $CP_1$ and $CP_2$. This is the first angular
point in the $T-U$ relation at $U\sim3.65$ in Fig.~\ref{UvT}(b).
The second angular point at $U\sim 4$ is the continuous
transition, connecting $CP_2$ to $HP$. The transition at lower
energy associated to the vanishing of $M$ is continuous in the
microcanonical ensemble with a negative specific heat, while
discontinuous in the canonical ensemble;
the dash-dotted line indicates the transition temperature in
the canonical ensemble,
derived using the Maxwell construction). The second transition
associated to the vanishing of $P$ is continuous in both
ensembles.

\section{Dynamical Properties I: Out-of-equilibrium states}
\index{non!equilibrium phenomena}

\subsection{Metastable states}\index{metastability}
Around the critical energy, relaxation to equilibrium depends in a
very sensitive way on the initial conditions adopted. When one starts
with out-of-equilibrium initial conditions in the ferromagnetic case,
one finds quasi-stationary (i.e. long lived) nonequilibrium
states\index{coherent!structures}. An example is represented by the
so-called ``water bag'' initial condition: all the particles are clustered
in a single point and the momenta are distributed according to a flat
distribution of finite width centered around zero.  These states have
a lifetime which increases with the number of particles~$N$, and are
therefore stationary in the continuum limit. In correspondence of
these metastable states, anomalous diffusion and L{\'e}vy walks
\cite{levyhmf}, long living correlations in $\mu-$space \cite{lat0} and
zero Lyapunov exponents \cite{lat0a} have been found.  In addition,
these states are far from the equilibrium caloric curve around the
critical energy, showing a region of negative specific heat and a
continuation of the high temperature phase (linear $T$ vs $U$
relation) into the low temperature one. It is very intriguing that
these out-of-equilibrium quasi-stationary states indicate a caloric
curve very similar to the one found in the region where one gets a
canonical first order phase transitions, but a continuous
microcanonical one, as discussed in section~\ref{microens}. In the
latter case, however, the corresponding states are stationary also at
finite $N$. The coexistence of different states in the continuum limit
near the critical region is a purely microcanonical effect, and arises
after the inversion of the $t \to \infty$ limit with the $N \to \infty$ one
\cite{lat0,lat0a,tsallisrap}.

Similarly, the antiferromagnetic HMF ($A=0$, $\varepsilon =-1$), 
where the particles interact through {\em repulsive} forces presents
unexpected dynamical properties in the
out-of-equili\-brium thermodynamics. On the first sight, the
thermodynamics of this model seems to be less interesting since no phase
transition occurs as discussed above. However, thermodynamical
predictions are again in some cases in complete disagreement with
dynamical results leading in particular to a striking localization
of energy. This aspect, as we will show below, is of course,
again, closely related to the long-range character of the
interaction and to the fact that such a dynamics is chaotic and
self-consistent. We mean by this that all particles give a contribution to the field
acting on each of them. One calls this phenomenon, 
{\em self-consistent chaos}~\cite{Diego}. In addition to the toy model that we
consider here, we do think that similar emergence of
structures, but even more importantly, similar dynamical
stabilization of out-equilibrium states could be encountered in
other long-range systems, as we briefly describe at the end of this
section.

\subsection{The dynamical emergence of the bicluster in the antiferromagnetic case}
\label{bicluster}

In the antiferromagnetic case, Eq.~(\ref{model0}) with $\varepsilon =-1$, the
intriguing properties appear in the region of very small energies. To
be more specific, if an initial state with particles evenly
distributed on the circle (i.e.  close to the ground state predicted
by microcanonical or canonical thermodynamics) and with vanishingly
small momenta is prepared, this initial condition can lead to the
formation of unstable states. This process, discovered by chance, is
now characterized in full
detail~\cite{Antoni,barreepjB,drh,firpoleyvraz}.

\begin{figure}[ht]
\begin{center}
\includegraphics[width=0.7\textwidth]{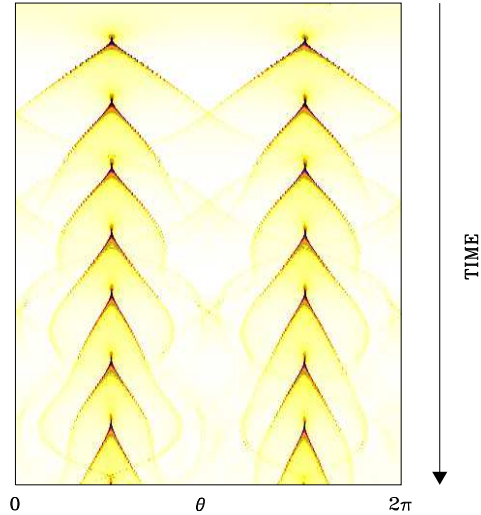}
\end{center}
\vskip-1truecm
 \caption[]{Short-time evolution of the particle density
in grey scale: the darker the grey, the higher the density.
Starting from an initial condition with all the particles evenly
distributed on the circle, one observes a very rapid concentration
of particles, followed by the quasi periodic appearance of
{\em chevrons}, that shrink as time increases.} \label{instab}
\end{figure}

As shown by Fig.~\ref{instab}, the density of particles is initially
homogeneous. However a localization of particles do appear at a given
time, in two different points, symmetrically located with respect to
the center of the circle. This localized state, that we call
{\it bicluster}\index{coherent!structures}, is however unstable (as shown
again by Fig.~\ref{instab}), since both clusters are giving rise to
two smaller localized groups of particles: this is the reason for the
appearance of the first {\em chevron}. However, also this state is
unstable, so that the first chevron disappears to give rise again to a
localization of energy in two points. This state enhances the
formation of a chevron with a smaller width and this phenomenon
repeats until the width of the chevron is so small that one does not
distinguish anymore its destabilization.
Asymptotically one gets a density distribution displaying two sharp
peaks located at distance $\pi$ on the circle, a
dynamically stable bicluster.

As we have shown in Ref.~\cite{barreepjB}, the emergence of the
bicluster is the signature of shock waves present in the
associated hydrodynamical equations. Indeed, we found a strikingly
good description of the dynamics of the particles by a non linear
analysis of the associated Vlasov equation, which is mathematically
justified~\cite{BraunHepp} in the infinite $N$ limit. 
The physical explanation of this problem can
be summarized as follows. Once the Hamiltonian has been mapped to
the Vlasov equation, it is possible to introduce a density $\rho(\theta,t)$ and a
velocity field $v(\theta,t)$. Neglecting the dispersion in
momentum and relying on usual non linear dynamics hierarchy of
time-scales, one ends up with dynamical equations at different 
orders in a multiscale analysis. The
first order corresponds to the linear dynamics and defines the
plasma frequency of order one. However, a second timescale appears
that is related to the previous one by the relationship
$\tau=\sqrt{U}\, t$, where $U$ is the energy per particle. When one
considers initial conditions with a very small energy density,
the two time scales are very different and  clearly
distinguishable by considering particle trajectories: a typical trajectory
corresponds indeed to a very fast motion with a very small
amplitude, superimposed to a slow motion with a large amplitude.

This suggests to average over the fast oscillations and leads to the
spatially forced Burgers equation
\begin{eqnarray}
\label{burgers} \frac{\partial u}{\partial \tau} +u\frac{\partial
u} {\partial \theta} & = & -\frac{1}{2}\sin 2\theta \qquad ,
\end{eqnarray}
once the average velocity $u(\theta,\tau)=\langle
v(\theta,t,\tau)\rangle_t$ is introduced. Due to the absence of dissipative
or diffusive terms,  Equation~(\ref{burgers}) supports shock
waves and these can be related to the emergence of the bicluster.
By applying the methods of characteristics to solve
Equation~(\ref{burgers}), one obtains
\begin{eqnarray}
\frac{d^2\theta}{d\tau^2} + \frac{1}{2} \sin 2\theta & = & 0\;.
\label{burgerslag}
\end{eqnarray}
which is a pendulum-like equation.

\begin{figure}[ht]
\begin{center}
\includegraphics[width=0.6\textwidth]{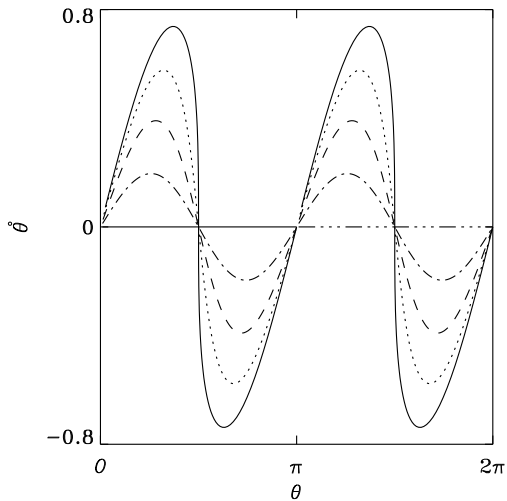}
\end{center}
 \caption[]{Five successive snapshots of the velocity
   profiles $u(\theta,t)$ are shown including the initial state when all
   particles are uniformly
   distributed in space with a small velocity dispersion.}
 \label{schock}
\end{figure}

Fig.~\ref{schock} shows the trajectories, derived from
Eq.~(\ref{burgerslag}), for particles that are initially evenly 
distributed on the circle. One clearly sees that two shock waves appear and
lead to an increase of the number of particles around two
particular sites, which depend on the initial conditions: this two
sites correspond to the nucleation sites of the bicluster. Because
of the absence of a diffusive term, the shock wave starts a spiral
motion that explains the destabilization of the first bicluster
and also the existence of the two arms per cluster, i.e. the
chevron.

\begin{figure}[ht]
\begin{center}
\includegraphics[width=0.6\textwidth]{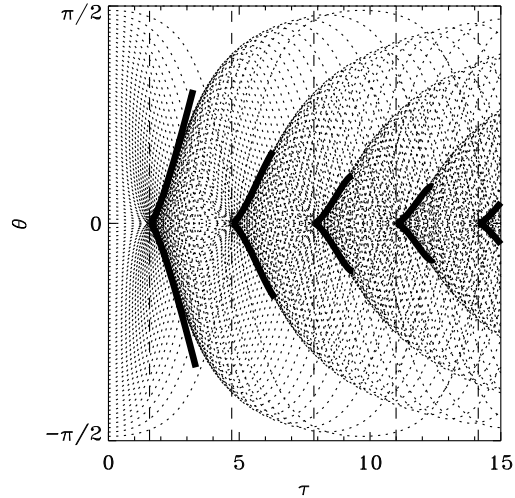}
\end{center}
\caption[]{Superposition of the caustics (thick full lines) over the 
characteristics (dotted lines) of
particles that are initially evenly distributed 
between $-\pi/2$ and $\pi/2$.}
\label{chevronsburgers}
\end{figure}

This dynamical analysis allows an even more precise description,
since the methods of characteristics show that the
trajectories correspond to the motion of particles in the
double well periodic potential $V(\theta,\tau)=-(\cos 2\theta)/4$. This
potential is of mean field origin, since it represents the effect
of all interacting particles. Therefore, the particles
will have an oscillatory motion in one of the two wells.
One understands thus that particles starting close to
the minimum will collapse at the same time, whereas a particle
starting farther will have a larger oscillation period. This is
what is shown in Fig.~\ref{chevronsburgers} where
trajectories are presented for different starting positions. One
sees that the period of recurrence of the chevrons corresponds to
half the period of oscillations of the particles close to the minimum;
this fact is a direct consequence of the isochronism of the
approximate harmonic potential close to the minimum. The
description could even go one step further by computing the
caustics, corresponding to the envelops of the characteristics.
They are shown in Fig.~\ref{chevronsburgers} and testify the
striking agreement between this description and the real
trajectories: the chevrons of Fig.~\ref{instab} correspond to the
caustics.

\subsection{Thermodynamical predictions versus dynamical stabilization}
\label{thermo}

If the above description is shown to be particularly accurate, it
does not explain why this state is thermodynamically preferred
over others. Indeed,
as shown before, thermodynamics predicts that the only
equilibrium state is homogeneous. This result has
been discussed in Section 3.1, where we have proved
that magnetization is zero at all energies for the antiferromagnetic
model and this has been also confirmed by Monte Carlo
simulations~\cite{drh}. However, since MD simulations
are performed at constant energy, it is important to derive
analytically the most probable state in the microcanonical
ensemble. Since this  model does not present ensemble inequivalence,
we can obtain the microcanonical results by employing the inverse 
Laplace transform (\ref{invLapla}) of the canonical partition function.
The microcanonical solution confirms that the maximal entropy state is 
homogeneous on the circle. 
It is therefore essential to see why the
bicluster state, predicted to be thermodynamically unstable is 
instead {\it dynamically stable}.

The underlying reason rely on the existence of the two very
different timesca\-les and the idea is again to average over the
very fast one. Instead of using the classical asymptotic expansion
on the equation of motions, it is much more appropriate to develop
an adiabatic approximation which leads to an effective Hamiltonian
that describes very well the long time dynamics. Doing statistical mechanics 
of this averaged problem, one predicts the presence of the bicluter.

The theory that we have developed relies on an application
of adiabatic theory, which in the case of the
HMF model is rather elaborate and needs lengthy
calculations~\cite{barreepjB} that we will not present here. 
An alternative, but less powerful, method to derive similar
results has been used in Ref.~\cite{firpoleyvraz}. 
On the contrary, we would like to present a qualitative explanations
of this phenomenon, using a nice (but even too simple !) analogy.

\begin{figure}[ht]
\begin{center}
\includegraphics[width=0.45\textwidth]{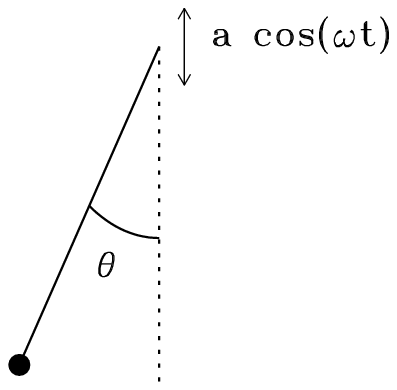}
\includegraphics[width=0.45\textwidth]{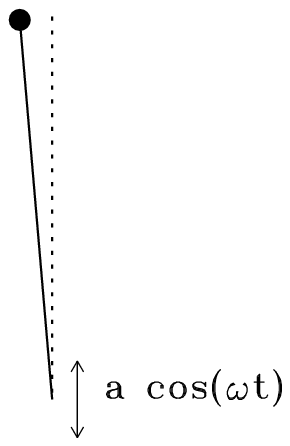}
\end{center}
\vskip-1truecm
 \caption[]{Schematic picture of the inverted pendulum.} \label{excitationparametrique}
\end{figure}

This stabilization of unstable states can be described using the
analogy with the inverted pendulum, where the vertical unstable
equilibrium position can be made stable by the application of
a small oscillating force. One considers a rigid rod
free to rotate in a vertical plane and whose point-of-support is
vibrated vertically as shown by
Fig.~\ref{excitationparametrique}a. If the support oscillates
vertically above a certain frequency, one discovers the remarkable
property that the vertical position with the center of mass above
its support point is {\em stable}
(Fig.~\ref{excitationparametrique}b). This problem, discussed
initially by Kapitza~\cite{Kapitaza}, has strong similarities with
the present problem and allows a very simplified presentation of
the averaging technique we have used.

The equation of motion of the vibrating pendulum is
\begin{equation}\label{eqpendukuminverted}
  \frac{d^2 \theta}{dt^2}+\left(\frac{\omega_0^2}{\omega^2}-a\cos t\right)\sin
  \theta=0~,
\end{equation}
where $\omega_0$ is the proper linear frequency of the pendulum, $\omega$ the
driving frequency of the support and  $a$ the amplitude of
excitation. Introducing a small parameter
$\varepsilon=\omega_0/\omega$, one sees
that~(\ref{eqpendukuminverted}) derives from the Lagrangian
\begin{equation}\label{Lagrangianpendulum}
  {\cal L}=\frac{1}{2}\left(\frac{d \theta}{d t}\right)^2+\left(\varepsilon^2-a\cos
  t\right)\cos\theta\quad.
\end{equation}
Here the two frequencies $\omega$ and $\omega_0$ define two
different time scales, in close analogy with the HMF model. Using
the small parameter to renormalize the amplitude of the excitation as
$A=a/\varepsilon$, and choosing the ansatz
$\theta=\theta_0(\tau)+\varepsilon\,\theta_1(t,\varepsilon)$ where
$\tau=\varepsilon t $, the Lagrangian equations for the function
$\theta_1$, leads to the solution $\theta_1=-A\cos
t\sin\theta_0(\tau)$. This result not only simplifies the above
ansatz, but more importantly suggest to average the Lagrangian on
the fast variable $t$ to obtain an effective Lagrangian $ {\cal
L}_{eff}=\langle {\cal
L}\rangle_t=\frac{1}{2}\left(\frac{d\theta_0}{d\tau}\right)^2+V_{eff}$~,
where the averaged potential is found to be
\begin{equation}\label{poteff}
V_{eff}=-\cos\theta_0-\frac{A^2}{8}\cos2\theta_0+\mbox{Cste}\quad.
\end{equation}
It is now straightforward to show that the inverted position would
be stable if $A^2>2$, i.e. if $a\omega>\sqrt{2}\omega_0$. As the
excitation amplitude $a$ is usually small, this condition
emphasizes that the two time scales should be clearly different,
for the inverted position to be stable.

The procedure for the HMF model is analogous, but of
course it implies a series of tedious calculations.
Since we would like to limit here to a pedagogical presentation,
we will skip such details that can be found in
Ref.~\cite{barreepjB}.
It is however important to emphasize
that the potential energy in the HMF model is self-consistently
determined and depend on the position of all particles. The magic
and the beauty is that, even if this is the potential energy of
$N$ particles, it is possible to compute the statistical mechanics
of the new effective Hamiltonian, derived directly from the
effective Lagrangian via the Legendre transform. The main result
is that the out-of-equilibrium state, (i.e. the bicluster
shown in Fig.~\ref{instab}) corresponds to a statistical
equilibrium of the effective mean-field dynamics.
No external drive is present in this case, as for the inverted pendulum, but
the time dependence of the mean field plays the role of the
external drive.

The HMF model represents presumably the simplest $N$-body
system where out-of-equilibrium dynamically stabilized states
can be observed and explained in detail.
However, we believe that several systems with long range interactions
should exhibit behaviours similar to the ones we have observed here.
Moreover, this model represents a paradigmatic example for
other systems exhibiting nonlinear interactions of rapid oscillations and
a slower global motion. One of this is the piston problem~\cite{piston}:
averaging techniques have been applied to the fast motion of gas particles in a
piston which itself has a slow motion~\cite{sinairusse}. Examples
can also be found in applied physics as for instance
wave-particles interaction in plasma physics~\cite{plasma}, or the
interaction of fast inertia gravity waves with the vortical motion
for the rotating Shallow Water model~\cite{embidmajda}.

\section{Dynamical Properties II: Lyapunov exponents}

In this section, we discuss the chaotic features of the microscopic
dynamics of the HMF model. We mainly concentrate on the $A=0$ case,
presenting in details the behavior of the Lyapunov exponents and the
Kolmogorov-Sinai entropy both for ferromagnetic and antiferromagnetic
interactions. We also briefly discuss some peculiar mechanisms of chaos in the $A
\neq 0$ case. The original motivation for the study of the chaotic
properties of the HMF was to investigate the relation between phase
transitions, which are macroscopic phenomena, and microscopic
dynamics (see Ref.~\cite{laporeport} for a review) with the purpose 
of finding dynamical
signatures of phase transitions\index{phase!transition}
\cite{cmd}. Moreover, we wanted to check the scaling properties with
the number of particles of the Lyapunov spectrum~\cite{livipolruffo} in 
the presence of long-range interactions.

\subsection{The $A=0$ case}

\begin{figure}[ht]
\begin{center}
\includegraphics[width=.7\textwidth]{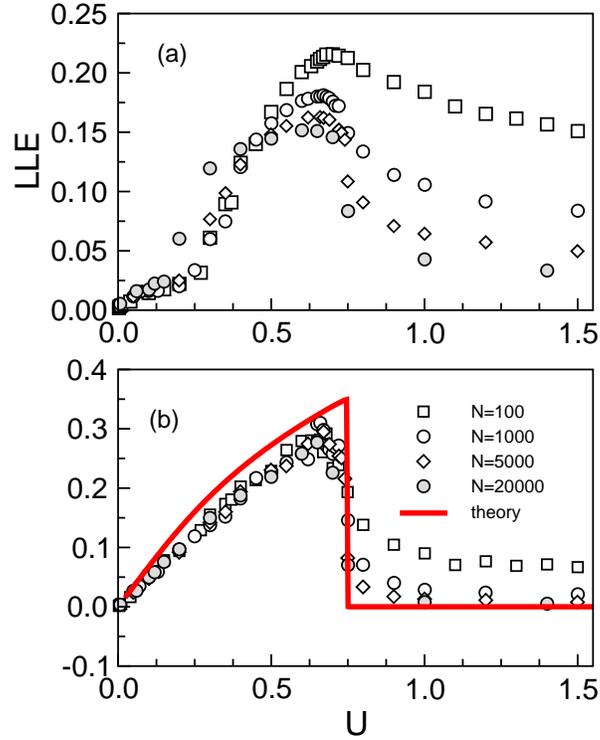}
\end{center}
\caption[] {Largest Lyapunov exponent LLE and kinetic energy
fluctuations $\Sigma= \sigma_K / \sqrt{N}$ as a function of U
in the $A=0$ ferromagnetic case for different N sizes. The theoretical curve
is shown as a full line, see text.}
\label{llehmf}
\end{figure}

In the $A=0$ case (see formulae (\ref{model0})-(\ref{pend})), the Hamiltonian 
equations of motion are
\begin{eqnarray}
\dot \theta_i&=&p_i\\
 \dot p_i  &=&-M \sin\left(\theta_i - \phi\right)\quad.\label{eqmoto}
\end{eqnarray}
The Largest Lyapunov Exponent (LLE) is defined as the limit
\begin{equation}
{\lambda_1}   = \lim_{t\to \infty}  {1\over t} \ln { d(t)\over
d(0) }
\end{equation}
where $d(t) =  \sqrt{ \sum_{i=1}^N  (\delta \theta_i)^2  + (\delta
p_i)^2 } $ is the Euclidean norm of the infinitesimal disturbance
at time $t$.  Therefore, in order to obtain the time evolution of
$d(t)$, one has to integrate also the linearized equations of
motion along the reference orbit
\begin{equation}
\label{lin1} {d\over dt} \delta \theta_i  =   \delta p_i ~~~~,~~~~
{d\over dt} \delta p_i ~= -\sum_j {{\partial^2V}\over {\partial
\theta_i \partial \theta_j} } \delta q_j~~,
\end{equation}
\noindent where the diagonal and off-diagonal terms of the Hessian are
\begin{eqnarray}
\label{diag} {\partial^2V\over{\partial \theta_i^2}} &=&M
\cos(\theta_i-\Phi) - \frac{1}{N}\\
 \label{der2}
{{\partial^2V}\over {\partial \theta_i \partial \theta_j} } &=&
-{1\over{N}}\cos(\theta_i-\theta_j)\quad,    ~~~i \neq j ~~.
\end{eqnarray}
To calculate the largest Lyapunov exponent we have used the
standard method by Benettin et al~\cite{ben}. In Fig.~(\ref{llehmf}),
we report the results obtained for four different sizes of the
system (ranging from $N=100$ to $N=20,000$).

In panel (a), we plot the largest Lyapunov exponent as a function
of U. As expected, $\lambda_1$ vanishes in the limit of very small
and very large energies, where the system is quasi-integrable.
Indeed, the Hamiltonian reduces to weakly coupled harmonic
oscillators in the former case or to free rotators in the latter.
For $U < 0.2$, $\lambda_1$ is small and has no $N$-dependence.
Then it changes abruptly and a region of ``strong chaos'' begins.
It was observed~\cite{Antoni} that between $U=0.2$ and $U=0.3$, a
different dynamical regime sets in and particles start to
evaporate from the main cluster, in analogy with what was
reported in other models~\cite{cmd,nayak}. In the region of strong
chaoticity, we observe a pronounced peak already for
$N=100$~\cite{yama}. The peak persists and becomes broader for
$N=20,000$. The location of the peak is slightly below the
critical energy and depends weakly on~$N$.

\begin{figure}[ht]
\begin{center}
\includegraphics[width=.7\textwidth]{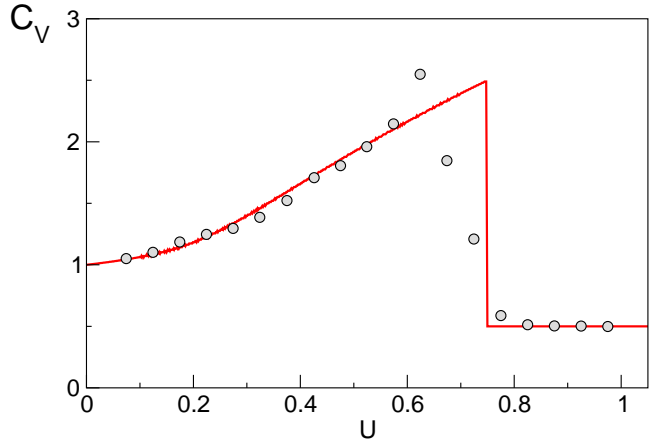}
\end{center}
\vskip 0.25truecm
\caption[]{Specific heat as a function of U
in the $A=0$ ferromagnetic case.
The numerical simulation at equilibrium for a system
with $N=500$ is compared with the
expected theoretical result (\protect\ref{cv1}).}
\label{cvhmf}
\end{figure}

In panel (b), we report the standard deviation of the kinetic
energy per particle $\Sigma$ computed from
\begin{equation}
\Sigma= \frac{\sigma_{K}}{\sqrt{N}}=
{{\sqrt  {\langle K^2\rangle -\langle K\rangle^2}} \over \sqrt{N}}\quad, \label{snum}
\end{equation}
where $\langle \bullet\rangle $ indicates the time average. The
theoretical prediction for $\Sigma$, which is also reported
in Fig.\ref{llehmf}, is~\cite{firpo,lat2hmf}
\begin{equation}
\Sigma= {T\over \sqrt{2}} \sqrt{  1 -{ \left[1-2 M \left({ dM\over
dT }\right)  \right]}^{-1} }\quad, \label{steo}
\end{equation}
where $M(T)$ is computed in the canonical ensemble. Finite size 
effects are also
present for the kinetic energy fluctuations, especially for
$U>U_c$, but in general there is a good agreement with the
theoretical formula, although the experimental points in
Fig.~\ref{llehmf}b lies systematically below it. The figure 
emphasizes that the
behavior of the Lyapunov exponent is strikingly correlated with
$\Sigma$: in correspondence to the peak in the LLE, we observe
also a sharp maximum of the kinetic energy fluctuations. The
relation between the chaotic properties and the thermodynamics of
the system, namely the presence of a critical point, can be made even more
quantitative. An analytical formula, relating (in the $A=0$ model)
the LLE to the second order phase transition undergone by the
system, has been obtained~\cite{firpo} by means of the
geometrical approach developed in Refs.~\cite{lapo1,laporeport}.
Using a reformulation of Hamiltonian dynamics in the language of
Riemannian geometry, they  have found a general analytical
expression for the LLE of a Hamiltonian many-body system in terms
of two quantities: the average $\Omega_0$ and the variance
$\sigma_{\Omega}$ of the Ricci curvature $ \kappa_R = \Delta V/N =
1/N~ \sum_{i=1}^N {\partial^2V\over{\partial q_i^2}} $, where $V$
is the potential energy and $q_i$ are the coordinates of the
system. Since in the particular case of the HMF model, we have
  \begin{equation}
{1\over N} \sum_{i=1}^N  {\partial^2V\over{\partial \theta_i^2}}
         =  M^2 -{1\over N}     = {2K\over N} + 1 - 2U - {1\over N} ~~,
\end{equation}
 the two quantities
$\Omega_0$ and $\sigma_{\Omega}$ can be expressed in terms of
average values and  fluctuations either of $M^2$ or of the kinetic
energy $K$
\begin{eqnarray}
\label{omegazerohmf} \Omega_0          &=&  \langle M^2\rangle -
\frac{1}{N}
                   =     \frac{2}{N}\langle K\rangle  +   (1-2U)  -\frac{1}{N}
\nonumber\\
\sigma^2_{\Omega} &=&           N  {\sigma^2_{M^2}}
                   =     \frac{4}{N} {\sigma^2_{K}~~.}
\end{eqnarray}
The formula obtained by Firpo~\cite{firpo} relates
the LLE, a characteristic   dynamical quantity, to thermodynamical quantities
like  $\langle M^2\rangle $ and ${\sigma_{M^2}}$,
or $\langle K\rangle $   and     ${\sigma_{K}}$, which
characterize the macroscopic  phase transition.
For moderately small values of $U$, an approximation of the formula
gives
\begin{equation}
\label{formulaprl}
\lambda
\propto \frac{\sigma_{K}}{\sqrt{N}}=\Sigma~~.
\end{equation}
This is in agreement with the proportionality between LLE and
fluctuations of the kinetic energy found numerically in
Fig.~(\ref{llehmf}). This implies also a connection between the LLE
and the specific heat, another quantity which is directly related
to the kinetic energy fluctuation. In fact the specific heat can
be obtained from $\Sigma$ by means of the Lebowitz-Percus-Verlet
formula~\cite{lpv}
\begin{equation}
C_V = { 1\over 2} \left[ 1- 2 \left({ \Sigma \over T} \right)^2
\right] ^{ -1} \quad. \label{cv1}
\end{equation}
In Fig.~\ref{cvhmf}, we report the numerical results for 
the specific heat as a function of $U$ for a system made of $N=500$
particles, and we compare them with the theoretical estimate.

In the HMF model and for a rather moderate size of the systems, it is
possible to calculate not only the LLE but all the Lyapunov exponents,
and from them the Kolmogorov-Sinai entropy.
We give first a succinct definition of the spectrum
of Lyapunov exponents (for more details see~\cite{Ruelleeckman}).
Once the 2N-dimensional tangent vector
${\bf z}=(\delta\theta_1,\cdots,\delta\theta_N,\delta p_1,\cdots,\delta p_N)$
is defined, with its dynamics given by Eqs.~(\ref{lin1}), one can formally
integrate the motion in tangent space up to time $t$, since the
equations are linear,
\begin{equation}
{\bf z}(t)=J^t{\bf z}(0)~,
\label{jey}
\end{equation}
where $J^t$ is a $2N\times2N$ matrix that depends on time
through the orbit $\theta_i(t),p_i(t)$.
The first $k$ exponents of the spectrum $\lambda_1,\dots,\lambda_k$,
which are ordered from the maximal to the minimal, are then given by
\begin{equation}
(\lambda_1 + \dots + \lambda_k) =
\lim_{t \to \infty} \frac{1}{2t}
\ln  \mbox{Tr} J_k^t (J_k^t)^*  \quad,
\label{Lyapk}
\end{equation}
where $J_k^t$ is the matrix ($(J_k^t)^*$ its transpose) 
induced by $J^t$ that acts
on the exterior product of $k$ vectors in the tangent space ${\bf
z}_1 \land \dots \land {\bf z}_k$. The spectrum
extends up to $k=2N$ and in our Hamiltonian system obeys
the pairing rule
\begin{equation}
\lambda_i = - \lambda_{2N + 1 - i} \quad \mbox{for} \quad 1 \leq i \leq 2N~.
\label{Pairing}
\end{equation}

\begin{figure}[ht]
\begin{center}
\includegraphics[width=.7\textwidth]{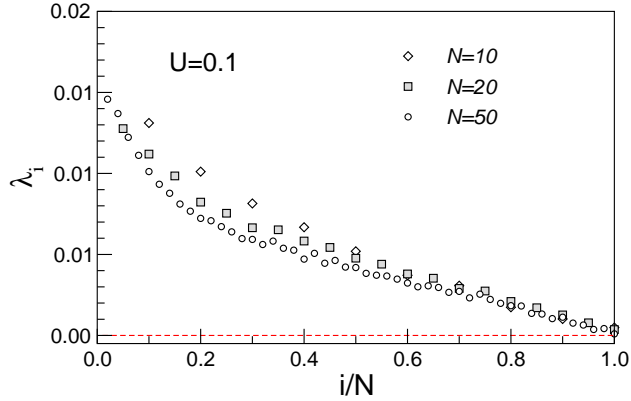}
\end{center}
\caption[]{Scaling of the positive part of the spectrum of Lyapunov
exponents in the A=0 ferromagnetic case for $U=0.1$.}
\label{spettro_f}
\end{figure}

The numerical evaluation of the spectrum of the Lyapunov exponents
is a heavy computational task, in particular for the necessity to
perform Gram-Schmidt orthonormalizations of the Lyapunov
eigenvectors in order to maintain them mutually orthogonal during
the time evolution.  We have been able to compute the complete
Lyapunov spectrum for system sizes up to $N=100$. In
Fig.~\ref{spettro_f}, we report the positive part of the spectrum
for different system sizes and an energy $U=0.1$ inside the weakly
chaotic region. The negative part of the spectrum is symmetric due
to the pairing rule (\ref{Pairing}). The limit distribution
$\lambda(x)$, suggested for short range interactions,
\begin{equation}
\lambda(x) = \lim_{N \to \infty} \lambda_{xN}(N)~,
\label{Distr}
\end{equation}
that is obtained by plotting $\lambda_i$ vs. $i/N$ and letting
$N$ going to infinity, is found also here for the
$N$ values that we have been able to explore.
At higher energies, this scaling is not valid and a size-dependence
is present~\cite{lat2hmf}.

\begin{figure}[ht]
\begin{center}
\includegraphics[width=.7\textwidth]{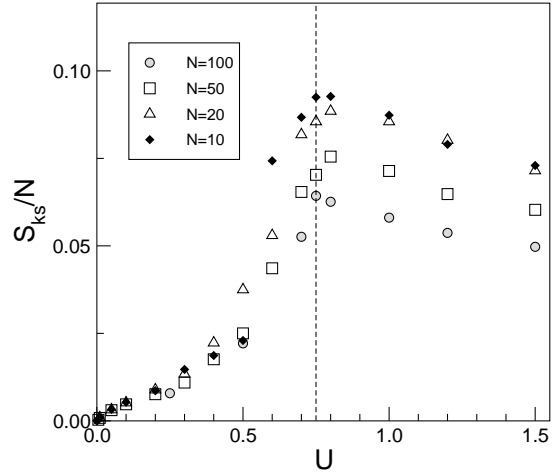}
\end{center}
\caption[]{$S_{KS}/N$ as a function of $U$
in the $A=0$ ferromagnetic case.Numerical calculations for different
systems sizes ranging from $N=10$ to $N=100$ are shown. The dashed
line indicates the critical energy.}
\label{sks_f}
\end{figure}

The Kolmogorov-Sinai (K-S) entropy is, according to Pesin's
formula~\cite{Ruelleeckman}, the sum of the positive Lyapunov
exponents. In Fig.~\ref{sks_f}, we plot the entropy density
$S_{KS}/N$ as a function of $U$ for different systems sizes.

\begin{figure}[ht]
\begin{center}
\includegraphics[width=.7\textwidth]{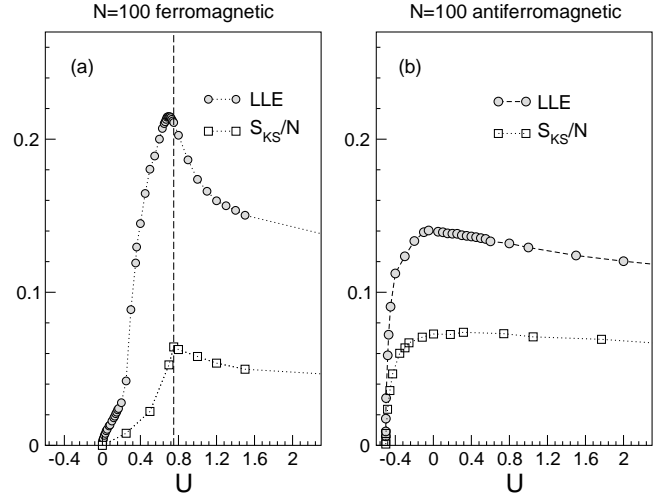}
\end{center}
\vskip0.5truecm
\caption[]{LLE and $S_{KS}/N$ as a function of $U$
in the $A=0$ ferromagnetic and
antiferromagnetic cases for N=100.}
\label{lle_fa}
\end{figure}

As for the LLE, $S_{KS}/N$ shows a peak near the critical energy,
a fast convergence to a limiting value as $N$ increases in the small
energy limit, and a slow convergence to zero
for $U \geq U_c$.
A comparison of the ferromagnetic and
antiferromagnetic cases is reported in Fig.~\ref{lle_fa}.
Here, for N=100, we plot as a function of $U$ the LLE and the
Kolmogorov-Sinai entropy per particle $S_{KS}/N$. In both the ferromagnetic
and antiferromagnetic cases, the system is integrable in the limits of
small and large energies. The main difference between the
ferromagnetic and the antiferromagnetic model appears at intermediate
energies. In fact, although both cases are chaotic (LLE and $S_{KS}/N$
are positive), in the ferromagnetic system one observes a well defined
peak just below the critical energy, because the dynamics feels the
presence of the phase transition\index{phase!transition}. On the other
hand, a smoother curve is observed in the antiferromagnetic case.
\begin{figure}[hb]
\begin{center}
\includegraphics[width=.7\textwidth]{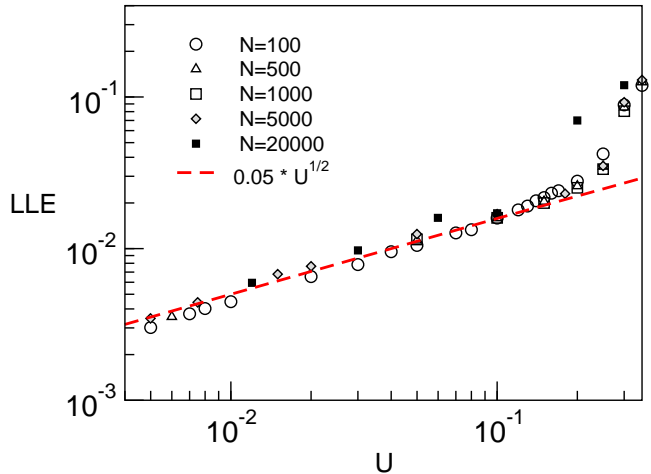}
\end{center}
\vskip0.5truecm
\caption[]{Scaling properties of the LLE at  low energies in the
$A=0$ ferromagnetic case. No N-dependence is observed for $U<0.2$.
The dashed line indicates a power-law $U^{1/2}$.}
\label{lle_low_f}
\end{figure}
In the low energy regime, it is possible to work out~\cite{lat1hmf} a
simple estimate $\lambda_1 \propto \sqrt{U}$, which is fully
confirmed for the ferromagnetic case in Fig.~\ref{lle_low_f} for
different system sizes in the range $N=100, 20000$. The same
scaling law is also valid in the antiferromagnetic
case~\cite{latoraprogtherophys} and for the $A=1$~\cite{at} HMF
model.

\begin{figure}[ht]
\begin{center}
\includegraphics[width=.9\textwidth]{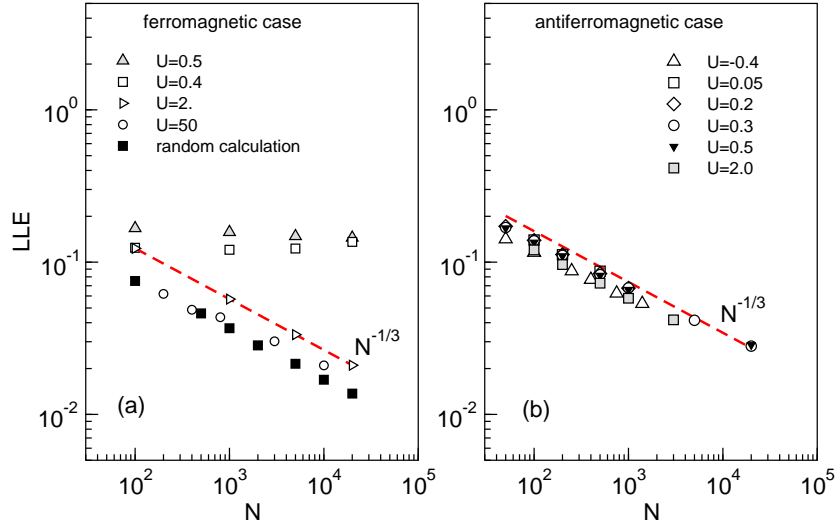}
\end{center}
\vskip1truecm
\caption{Scaling of the LLE vs $N$  for the $A=0$ ferromagnetic and
antiferromagnetic cases at various energies, see text. A power-law
decreasing as $N^{-1/3}$ of the LLE is observed for overcritical energies
in the ferromagnetic case and for all energies in the antiferromagnetic one.
See text for further details.}
\label{lle_high_fa}
\end{figure}

At variance with the $N$-independent behavior observed at small
energy, strong finite size effects are present above the critical
energy in the ferromagnetic case and for all energies for the
antiferromagnetic case. In Fig.~\ref{lle_high_fa}(a), we show that
the LLE is positive and $N$-independent below the transition (see
the values $U=0.4, 0.5$), while it goes to zero with  $N$ above.
We also report in the same figure a calculation of the LLE using a
random distribution of particle positions $\theta_i$ on the circle
in Eq.~(\ref{lin1}) for the tangent vector. The agreement between
the deterministic estimate and this random matrix calculation is
very good. The LLE scales as $N^{-{1\over3}}$, as indicated by the
fit reported in the figure. This agreement can be explained by means
of an analytical result obtained for the LLE of product of random
matrices~\cite{pari}. If the elements of the symplectic random
matrix have zero mean, the LLE scales with the power $2/3$ of the
perturbation. In our case, the latter condition is satisfied and
the perturbation is the magnetization $M$. Since $M$ scales as
$N^{-{1\over 2}}$, we get the right scaling of $\lambda_1$ with
$N$. This proves that the system is integrable for $U \geq U_c $
as $N \to \infty$. This result is also confirmed by the analytical
calculations of Ref.~\cite{firpo} and, more recently, of 
Ref.~\cite{Vallejos}.

In the antiferromagnetic case, the LLE goes to zero with 
system size as $N^{-{1\over 3}}$ for all values of $U$.

Interesting scaling laws have also been found for the
Kolmogorov-Sinai entropy in the ferromagnetic case: at small
energies $S_{KS}/N \propto U^{3/4}$ with no size dependence,  and
$S_{KS}/N \propto N^{-1/5}$ for overcritical energy densities. The
latter behavior has been found also in other
models~\cite{tsuchia}. Concluding this section we would like to
stress that the finite value of chaotic measures close to the
critical point is strongly related to kinetic energy fluctuations and can
be considered as a {\em microscopic dynamical indication} of the
macroscopic equilibrium phase transition. This connection has been
found also in other models and seems to be quite
general~\cite{laporeport,cmd,lapo3,barrehmf}

The behavior of the HMF model as a function of the range
of the interaction~\cite{celiahmf,tamahmf,campa} and the dynamical
features before equilibration \cite{lat0,lat0a} is discussed in a
separate chapter of this volume in connection to Tsallis
nonextensive thermodynamics \cite{tsallisrap}.

\subsection{Mechanisms of chaos in the $A\neq 0$ case}

For $A=0$, the origin of chaos is related to the non time-dependence 
of Eq.~(\ref{pend}), since it is obvious that if the
phase $\phi$ and the magnetization $M$ would become constant the
dynamics of the system will reduce to that of an integrable
system. There are indeed preliminary indications~\cite{priv} that in the
mean-field limit $N \to \infty$, $M$ and $\phi$ will become
constant and $\lambda \to 0$ . It should be noticed that this is
true if the mean-field limit is taken before the limit $t \to
\infty$ in the definition of the maximal Lyapunov exponent, and
numerical indications were reported in Ref.~\cite{lat0a}. When $A
> 0$ we expect a quite different situation: indeed, even
assuming that in the mean-field limit $M$ and $P$ and their
respective phases will become constant, the dynamics will
eventually take place in a $4$-dimensional phase space and chaos
can in principle be observed.

As already shown in~\cite{at}, for $A=1$ and $\varepsilon=1$, two
different mechanisms of chaos are present in the system for $U <
U_c$ : one acting on the particles trapped in the potential and
another one felt by the particles moving in proximity of the
separatrix. This second mechanism is well known and is related to
the presence of a chaotic layer situated around the separatrix.
The origin of the first mechanism is less clear, but presumably
related to the erratic motion of the minimum of the potential
well, i.e. to the time-dependent character of the equations ruling
the dynamics of the single particle. Indications in this direction
can be found by performing the following numerical experiment. Let
us prepare a system with $N=200$ and $U=0.87$ (the critical energy
is in this case $U_c \sim 2$) with a Maxwellian velocity
distribution and with all particles in a single
cluster.

\begin{figure}[ht]
\begin{center}
\includegraphics[width=0.5\textwidth,angle=270]{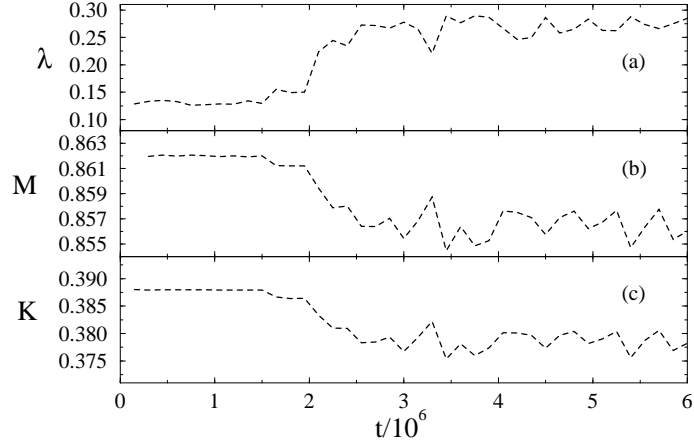}
\end{center}
\caption[]{Time evolution of the Largest Lyapunov Exponent (LLE)
$\lambda_1$, of the magnetization $M$, and of the kinetic
energy $K$ are shown for a clustered initial condition
for the model with $A=1$ and $\varepsilon=1$ at $U=0.87$ and with $N=200$.
}
\label{lyap}
\end{figure}

For an integration time $ t < 2 \times 10^6$, the Lyapunov
exponent has a value $\lambda \simeq 0.13$. But when at time $t
\sim 2 \times 10^6$, one particle escapes from the cluster, its
value almost doubles (see Fig.~\ref{lyap}). The escaping of the
particle from the cluster is associated to a decrease of the
magnetization $M$ and of the kinetic energy~$K$. This last effect
is related to the negative specific heat regime: the potential
energy $V_A$ is minimal when all the particles are trapped, if one
escapes then $V_A$ increases and due to the energy conservation
$K$ decreases. As a matter of fact, we can identify a ``strong''
chaos felt from the particles approaching the separatrix and a
``weak'' chaos associated to the orbits trapped in the potential
well. We believe that the latter mechanism of chaotization should
disappear (in analogy with the $A=0$ case) when the mean-field
limit is taken before the $t \to \infty$ limit. Therefore we
expect that for $N \to \infty$ the only source of chaotic
behaviour should be related to the chaotic sea located around the
separatrix. As already noticed in Ref.~\cite{at}, the degree of
chaotization of a given system depends strongly on the initial
condition (in particular in the mean-field limit). In the latter
limit, for initial condition prepared in a clustered
configuration, we expect that $\lambda =0$, until one particle will
escape from the cluster.

\section{Conclusions}

We have discussed the dynamical properties of the Hamiltonian Mean
Field model in connection to ita thermodynamics.  This apparently
simple class of models has revealed a very rich and interesting
variety of behaviours. Inequivalence of ensembles, negative specific
heat, metastable dynamical states and chaotic dynamics are only some
among them. During the past years these models have been of great
help in understanding the connection between dynamics and
thermodynamics when long-range interactions are present.  Such
kind of investigation is of extreme importance for self-gravitating
systems and plasmas, but also for phase transitions in finite systems,
such as atomic clusters or nuclei, and for the foundation of
statistical mechanics.  Several progresses have been done during these
years.  This contribution, although not exhaustive, is an
effort to summarize some of the main results achieved so far.  
We believe that the problems which are still not understood 
will be hopefully clarified in the near future within a general 
theoretical framework. We list three important open questions that
we believe can be reasonably addresses: the exact solution of the 
model in the microcanonical ensemble; the full characterization 
of the out-of-equilibrium states close to phase transitions; the 
clarification of the scaling laws of the maximal Lyapunov exponent 
and of the Lyapunov spectrum; the study of the single-particle 
diffusive motion~\cite{levyhmf} in the various non-equilibrium 
and equilibrium regimes.

\section*{Acknowledgements}
We would like to warmly thank our collaborators Mickael Antoni, Julien
Barr{\'e}, Freddy Bouchet, Marie-Christine Firpo, Fran{\c c}ois Leyvraz and
Constantino Tsallis for fruitful interactions. One of us (A.T.)  would
also thank Prof. Ing. P. Miraglino for giving him the opportunity to
complete this paper.  This work has been partially supported by the EU
contract No. HPRN-CT-1999-00163 (LOCNET network), the French
Minist{\`e}re de la Recherche grant ACI jeune chercheur-2001 N$^\circ$
21-311.  This work is also part of the contract COFIN00 on {\it Chaos
and localization in classical and quantum mechanics}.

\end{document}